\documentclass[11pt,a4paper]{article}
\usepackage{jheppub_kim}
\usepackage{rotating} 
 
\usepackage{graphicx,epsfig}
\usepackage{amsmath}
\usepackage {amssymb}
\usepackage{subfigure}
 \def\e{\mathrm{e}}

\usepackage{relsize}

\usepackage{graphicx}
\usepackage{epsfig}
\usepackage{bm}
\usepackage{amssymb}
\usepackage{float}
\usepackage{amsmath}
\usepackage{dcolumn}
\usepackage{cancel}

\providecommand{\U}[1]{\protect\rule{.1in}{.1in}}

\newcommand{\newc}{\newcommand}

\newc{\be}{\begin{equation}}
\newc{\ee}{\end{equation}}

\newc{\ba}{\begin{eqnarray}}
\newc{\ea}{\end{eqnarray}}
\newc{\bea}{\begin{eqnarray*}}
\newc{\eea}{\end{eqnarray*}}
\newc{\D}{\partial}
\newc{\ie}{{\it i.e.} }
\newc{\eg}{{\it e.g.} }
\newc{\etc}{{\it etc.} }
\newc{\etal}{{\it et al.}}
\newc{\lcdm}{$\Lambda$CDM }

\newc{\ra}{\Rightarrow}

\title{Kaniadakis holographic dark energy and cosmology}

\author[a]{Niki Drepanou}

\author[b]{Andreas Lymperis}

\author[c,d,e]{Emmanuel N. Saridakis}
 
 \author[e,f]{Kuralay Yesmakhanova}

\affiliation[a]{Department of Physics, National Technical University of Athens, 
Zografou
Campus GR 157 73, Athens, Greece}

\affiliation[b]{Department of Physics, University of Patras, 26500 Patras, 
Greece}

\affiliation[c]{National Observatory of Athens, Lofos Nymfon, 11852 Athens, 
Greece}

\affiliation[d]{CAS Key Laboratory for Researches in Galaxies and Cosmology, 
Department of Astronomy, University of Science and Technology of China, Hefei, 
Anhui 230026, P.R. China}

\affiliation[e]{Eurasian National University, Nur-Sultan Astana
010008, Kazakhstan}

 \affiliation[f]{Ratbay Myrzakulov Eurasian International Centre for Theoretical 
Physics, Nur-Sultan 010009, Kazakhstan}

\emailAdd{alymperis@upatras.gr}
\emailAdd{msaridak@phys.uoa.gr}

\abstract{ 
We construct a holographic dark energy scenario based on Kaniadakis  entropy, 
which is a generalization of Boltzmann-Gibbs entropy that arises from 
relativistic statistical theory and is characterized by a single parameter $K$ 
which 
quantifies the deviations from standard expressions, and we use the 
future event horizon as the Infrared cutoff. We extract the differential 
equation that determines the evolution of the effective dark energy density 
parameter, and we provide   analytical expressions for the corresponding 
equation-of-state  and deceleration parameters. We show that the universe 
exhibits the 
standard thermal history, with the sequence of matter and dark-energy eras, 
while the transition to acceleration takes place at $z\approx0.6$.
Concerning the dark-energy equation-of-state parameter we show that it can have 
a rich behavior, being quintessence-like, phantom-like, or experience the 
phantom-divide crossing in the past or in the future. Finally, in the far 
future dark energy dominates completely, and the asymptotic value of its  
equation of state depends on the values 
of the two model parameters.  
}

\begin{document}
\maketitle

\section{Introduction}

It is now well established that the Universe at late times  experienced the 
transition from the matter era to the accelerated expansion phase. Although the 
simplest explanation would be the consideration of the cosmological constant, 
the corresponding problem related to the quantum-field-theoretical calculation 
of its value, as well as the possibility of a dynamical nature, led to two main 
paths of constructing extended scenarios. The first is to maintain general 
relativity as the underlying theory of gravity, and consider  new, exotic forms 
of matter that constitute the concept of dark energy  
\cite{Copeland:2006wr,Cai:2009zp,Bamba:2012cp}. The second is to
construct extended or modified theories of gravity, that posses general 
relativity as a low-energy limit, but which in general provide the extra 
degrees of freedom that can drive the dynamical universe acceleration  
\cite{Nojiri:2010wj,Capozziello:2011et,Cai:2015emx,CANTATA:2021ktz}.
 
Nevertheless, one can acquire an alternative   explanation of the dark 
energy  origin, through the cosmological application 
\cite{Fischler:1998st,Bak:1999hd,Horava:2000tb} of the 
 holographic principle 
\cite{tHooft:1993dmi,Susskind:1994vu,Bousso:2002ju}. 
The corresponding  framework is based on the 
thermodynamics of black holes   and the connection   of the Ultraviolet
cutoff of a quantum field
theory (related 
to the vacuum energy), with the largest
distance of the theory (which in turn is a requirement in order for the theory 
to be applicable at large distances)  \cite{Cohen:1998zx}.
In particular, in a given system  whose
entropy is   proportional to its volume, the total energy    
should not be larger than the mass of a black hole with the same size, whose 
entropy is proportional to its area,
since in such a case the
system would collapse to a black hole. If one considers the whole Universe as 
the system  one  extracts    
 a vacuum energy  of holographic origin, namely a form of
  holographic dark energy with dynamical nature  
\cite{Li:2004rb,Wang:2016och}.

  The cosmological implications of holographic dark energy proves to be very 
interesting
\cite{Li:2004rb,Wang:2016och,Horvat:2004vn,Huang:2004ai,Pavon:2005yx,
Wang:2005jx,
Nojiri:2005pu,Kim:2005at,
Wang:2005ph, Setare:2006wh,Setare:2008pc,Setare:2008hm} and it proves to be 
 in agreement  with  observations
\cite{Zhang:2005hs,Li:2009bn,Feng:2007wn,Zhang:2009un,Lu:2009iv,
Micheletti:2009jy}. 
In this scenario one is free of the naturalness problem of the cosmological 
constant \cite{Weinberg:1988cp}, as well as from   pathologies that may 
arise 
in various modified gravity constructions  
\cite{CANTATA:2021ktz}. Additionally, it has been shown to be able to 
alleviate the $H_0$ and growth tensions between $\Lambda$CDM scenario and some 
direct measurements \cite{Abdalla:2022yfr}. That is why a large amount of 
research has been devoted to these  investigations, and 
the basic models  have been   extended 
in various ways 
\cite{Gong:2004fq,Saridakis:2007cy,  
Setare:2007we,Cai:2007us,Setare:2008bb,Saridakis:2007ns,Saridakis:2007wx,
Jamil:2009sq,
Gong:2009dc, 
Suwa:2009gm,Jamil:2010vr,BouhmadiLopez:2011xi,Malekjani:2012bw,
Khurshudyan:2014axa,
Landim:2015hqa,Pasqua:2015bfz,
Jawad:2016tne,Pourhassan:2017cba,Saridakis:2017rdo,Nojiri:2017opc,
Saridakis:2018unr,  
Oliveros:2019rnq,Kritpetch:2020vea,Saridakis:2020zol,Dabrowski:2020atl,
daSilva:2020bdc, Anagnostopoulos:2020ctz, Mamon:2020spa,
Bhattacharjee:2020ixg,Huang:2021zgj,Lin:2021bxv,Colgain:2021beg, 
Hossienkhani:2021emv,Nojiri:2021iko,Shekh:2021ule,Telali:2021jju,
Shaikh:2022ynt}.

The basic expression in the  construction of holographic dark energy 
is the one that connects the entropy of a system  with   geometrical 
quantities such as its radius. The standard one is the  Bekenstein-Hawking 
entropy, which arises as the black-hole and cosmological application of the 
standard Boltzmann-Gibbs entropy. However,  Kaniadakis has proposed   a 
one-parameter generalization of the   Boltzmann-Gibbs  entropy, called 
Kaniadakis entropy
 \cite{Kaniadakis:2002zz,Kaniadakis:2005zk}. This results from 
  a  
self-consistent and coherent relativistic statistical 
theory, in which the basic features of standard statistical theory are 
maintained. In such an extended statistical theory, the
  distribution functions are a one-parameter continuous deformation 
of the usual Maxwell-Boltzmann ones, and hence standard statistical theory is 
recovered in a particular limit.

In the present work we will use Kaniadakis entropy in order to formulate 
Kaniadakis holographic dark energy, and study its cosmological implications.
Although in the literature there were some first attempts towards this 
direction  \cite{Moradpour:2020dfm,Jawad:2021xsr,Sharma:2021zjx}
the resulting models were not correct.
The reason for this failure was  the fact that the authors used
the Hubble horizon instead of the future event horizon in the basic holographic 
expression. Therefore, not only the resulting models could not recover usual 
holographic dark energy in the limit where  Kaniadakis entropy becomes standard 
entropy, as it should, but in order to be able to describe the universe 
evolution one needs unacceptably large values of the Kaniadakis parameter, 
namely unacceptably large deviations from standard entropy.
 Hence, in this work we proceed to the consistent formulation
of Kaniadakis holographic dark energy, which is indeed a well-defined 
extension  of standard holographic dark energy,
recovering it as a particular limit in the case where 
Kaniadakis entropy becomes standard Bekenstein-Hawking 
entropy.
  
The plan of the manuscript is the following: In Section   \ref{model} we 
formulate Kaniadakis holographic dark energy,  we present the corresponding 
cosmological equations and we extract analytical relations for the dark energy 
density and equation-of-state parameters. Then in Section 
\ref{Cosmologicalevolution} we proceed to the  study of the resulting 
cosmological behavior. Finally, in Section \ref{Conclusions} we discuss our 
results and we summarize.

\section{Kaniadakis holographic dark energy}
\label{model}
  
  In this section we proceed to the formulation of Kaniadakis holographic 
dark energy. The basic idea behind  holographic dark energy is the 
 inequality $\rho_{DE} 
L^4\leq S$, where $L$ is the largest distance of 
the theory (the  Infrared  cutoff) and $S$ the entropy relation applied in a 
black hole of radius $L$  
\cite{Li:2004rb,Wang:2016och}. In the case of standard  
Bekenstein-Hawking 
entropy
$S_{BH}\propto A/(4G)=\pi L^2/G$, with $G$ the Newton's constant,  the 
saturation of the above inequality gives standard holographic dark energy, 
namely  $\rho_{DE}=3c^2 M_p^2 L^{-2}$, with   $M_p$ the Planck mass and $c$ the 
model parameter. Hence, we can see that if instead of standard entropy we 
use a modified one, we will obtain a modified holographic dark energy.
  
As we mentioned in the Introduction, Kaniadakis entropy   is a one-parameter
generalization of the classical   entropy. It is given 
by  \cite{Kaniadakis:2002zz,Kaniadakis:2005zk}
\begin{eqnarray}
S_{K}=- k_{_B} \sum_i n_i\, \ln_{_{\{{\scriptstyle
K}\}}}\!n_i  ,
\end{eqnarray}
with $k_{_B}$ the Boltzmann constant,  and
where we have defined $\ln_{_{\{{\scriptstyle
K}\}}}\!x=(x^{K}-x^{-K})/2K$. Kaniadakis entropy is characterized by the single 
dimensionless parameter $K$, which  quantifies the 
deviation from the case of standard statistical mechanics. Hence, standard 
entropy is recovered in the limit $K\rightarrow0$, while $K$ can vary in the 
range $-1<K<1$. Additionally, in such a 
 generalized statistical theory the distribution function reads 
as
$ n_i= \alpha \exp_{_{\{{\scriptstyle K}\}}}[-\beta
(E_i-\mu)]  , 
$
where $\exp_{_{\{{\scriptstyle K}\}}}(x)=
\left(\sqrt{1+K^2x^2}+K x\right)^{1/K}$, 
 $\alpha=[(1-K)/(1+K)]^{1/2K}$, 
$1/\beta=\sqrt{1-K^2}\,\,k_{_{B}}\!T$, and the chemical
potential $\mu$ can be fixed by normalization 
\cite{Kaniadakis:2002zz,Kaniadakis:2005zk}.  
 Kaniadakis entropy    can be expressed as 
\cite{Abreu:2016avj,Abreu:2017fhw,Abreu:2017hiy,Abreu:2018mti,Yang:2020ria,
Abreu:2021avp}
\begin{equation}
 \label{kstat}
S_{K} =-k_{_B}\sum^{W}_{i=1}\frac{P^{1+K}_{i}-P^{1-K}_{i}}{2K},
\end{equation}
where  $P_i$ is the probability the system to be in  a specific microstate and 
 $W$  the total number of configurations.
 
 Let us apply  Kaniadakis entropy   in the black-hole framework, which will 
then be needed for the holographic application.  Assuming that 
  $P_i=1/W$,     using the fact 
that Boltzmann-Gibbs entropy is $S\propto\ln(W)$, while   the 
Bekenstein-Hawking entropy is given by $S_{BH}= A/(4G)$,
  we acquire $W=\exp\left[ A/(4G)\right]$ \cite{Moradpour:2020dfm}, 
where from now on we impose      units in which the 
Boltzmann constant, the light speed, and the reduced Planck constant are set to 
 $k_{_B}=c=\hbar=1$.
Hence, inserting these into (\ref{kstat}) we find 
 \begin{equation} \label{kentropy}
S_{K} = \frac{1}{K}\sinh{(K S_{BH})}.
\end{equation}
As expected in the limit  $K\rightarrow 0$ one recovers   standard 
Bekenstein-Hawking entropy, 
i.e. $S_{K\rightarrow 0}=S_{BH}$. 
Since in reality one expects the above modified entropy to be close to the 
standard Bekenstein-Hawking value, we expect that $K\ll1$ (we remind that 
$-1<K<1$). Thus, 
it is justified to expand the above Kaniadakis entropy for small $K$, 
obtaining
\begin{equation}\label{kentropy2}
S_{K} = S_{BH}+ \frac{K^2}{6} S_{BH}^3+ {\cal{O}}(K^4).
\end{equation}
As one can see, the first term is the usual entropy, while the second term is 
the lowest-order Kaniadakis correction.

It is now easy to extract the relation of Kaniadakis  holographic dark energy. 
In particular, inserting (\ref{kentropy2}) into the inequality 
$\rho_{DE} L^4\leq S$, we obtain 
  \begin{equation}
\label{KHDE}
\rho_{DE}= 3c^2 M_p^2 L^{-2}+ 3\tilde{c}^2K^2 M_p^6 L^2,
\end{equation}
with $c$ and $\tilde{c}$  constants. As 
mentioned above, for 
$K=0$ the above expression gives the 
usual holographic dark energy  $\rho_{DE}=3c^2 M_p^2 L^{-2}$.  
 In the following we absorb the constant  $\tilde{c}$ inside the parameter $K$, 
by setting $3\tilde{c}^2K^2\equiv\tilde{K}^2$ and we drop the tildes for   
simplicity.

We proceed by considering   a flat homogeneous and isotropic 
Friedmann-Robertson-Walker (FRW) geometry with
  metric  
\begin{equation}
\label{FRWmetric}
ds^{2}=-dt^{2}+a^{2}(t)\delta_{ij}dx^{i}dx^{j}\,,
\end{equation}
with $a(t)$ the scale factor.   
As a next step, in  any holographic dark 
energy scenario, one needs 
to determine the   length   $L$ that appears in the corresponding relations. In 
the case of standard holographic dark energy models it is well 
known that  $L$ cannot be the Hubble 
horizon $H^{-1}$ (where $H\equiv \dot{a}/a$ is the Hubble function), since 
this choice leads to obvious
inconsistencies \cite{Hsu:2004ri}, such as no acceleration. Thus, one must use
 the   future event horizon 
  \cite{Li:2004rb} 
\begin{equation}
\label{futrhoriz}
R_h\equiv a\int_t^\infty \frac{dt}{a}= a\int_a^\infty \frac{da}{Ha^2}.
\end{equation}
As we mentioned in the Introduction, in some 
recent attempts to construct Kaniadakis holographic dark energy the authors 
used   (\ref{kentropy}) but then they considered  the Hubble horizon to be $L$ 
\cite{Moradpour:2020dfm,Jawad:2021xsr,Sharma:2021zjx}.  Thus, the obtained 
models do not have standard holographic dark energy and standard 
thermodynamics as  a sub-case, and this is a serious 
disadvantage.   One can  verify that in a clear way by observing 
that in order to have reasonable observational results the authors demand 
$K$ values of the order of $10^3$,  namely a huge 
deviation from standard Bekenstein-Hawking 
entropy, which is  not observed (not mentioning the fact that the 
initial $K$ parameter of Kaniadakis entropy is bounded in $-1<K<1$).

In the present work we desire to formulate  Kaniadakis 
holographic dark 
energy in a consistent way, and hence we use as $L$ the future event horizon 
(\ref{futrhoriz}). In this way, as we will see, standard holographic dark 
energy is included as a sub-case, and can be obtained for $K\rightarrow0$.
However, let us comment here that  using  the future event horizon  does 
have 
the disadvantage that   the dark-energy 
density  at present  depends on the future expansion of the Universe 
\cite{Wang:2016och}, in a similar way that  the use of the Hubble 
horizon has   the disadvantage that the dark-energy density, which is a 
local concept,   depends on the global picture of the whole Universe, namely 
on the global expansion scale. Nevertheless, we mention that the use of both 
horizons, as well as other horizons, such as the Granda-Oliveros  cutoff 
\cite{Granda:2008dk}, is in principle justified exactly by the concept of 
holography and the dualities 
known from string theory, that relate the very small with the very large in 
space and time.

According to the above discussion, and using (\ref{KHDE}) with $L$ the $R_h$, 
the energy 
density of Kaniadakis holographic dark energy writes as
 \begin{equation}
\label{KHDE2}
\rho_{DE}= 3c^2 M_p^2 R_h^{-2}+ K^2 M_p^6 R_h^2.
\end{equation} 
The Friedmann equations in a universe containing the dark energy and matter 
perfect fluids are
 \begin{eqnarray}
\label{Fr1b}
3M_p^2 H^2& =& \ \rho_m + \rho_{DE}    \\
\label{Fr2b}
-2 M_p^2\dot{H}& =& \rho_m +p_m+\rho_{DE}+p_{DE},
\end{eqnarray}
with $p_{DE}$ the pressure of  Kaniadakis holographic dark energy, and $\rho_m$ 
and $p_m$ 
respectively the energy density and pressure of the matter sector.
The equations close by considering the matter
conservation 
equation  
\begin{equation}\label{rhoconserv}
\dot{\rho}_m+3H(\rho_m+p_m)=0.
\end{equation}

It proves convenient to introduce the dark energy and matter density parameters 
through
 \begin{eqnarray}
 && \Omega_m\equiv\frac{1}{3M_p^2H^2}\rho_m
 \label{Omm}\\
 &&\Omega_{DE}\equiv\frac{1}{3M_p^2H^2}\rho_{DE}.
  \label{ODE}
 \end{eqnarray}
 Using these definitions, relations (\ref{futrhoriz}),(\ref{KHDE2}),(\ref{ODE}) 
lead to
  \begin{equation}\label{integrrelation}
\int_x^\infty \frac{dx}{Ha}=\frac{1}{a}\left( \frac{3H^2 \Omega_{DE}-
\sqrt{9H^4 \Omega_{DE}^2 -12c^2 K^2M_p^4} }{2K^2M_p^4}
\right)^{\frac{1}{2}},
\end{equation}
 where  $x\equiv \ln a$. Note that solving the fourth-degree algebraic equation 
(\ref{KHDE2}) we have kept only the solutions that give positive $R_h$ and 
moreover with the usual limiting result for  $K\rightarrow0$. Indeed, as one 
can see, in the limit $K\rightarrow0$ the above relation gives the 
standard holographic dark energy result
$
\int_x^\infty \frac{dx}{Ha}=
\frac{c}{aH\sqrt{\Omega_{DE}}}$.

We focus  on the physically interesting dust matter case, where
the 
matter 
equation-of-state parameter is set to zero. Therefore,
(\ref{rhoconserv})  
leads to $\rho_m=\rho_{m0}/a^3$, where $\rho_{m0}$ is the       matter 
energy 
density  
at the current scale factor $a_0=1$ (we use the subscript ``0'' to denote   
the  
present value of a quantity).  Hence, substituting   into (\ref{Omm}) leads to 
$\Omega_m=\Omega_{m0} H_0^2/(a^3 H^2)$, and then, using   the Friedmann 
equation    $\Omega_m+\Omega_{DE}=1$, 
we find
 \begin{equation}\label{Hrel2}
\frac{1}{Ha}=\frac{\sqrt{a(1-\Omega_{DE})}}{H_0\sqrt{\Omega_{m0}}}.
\end{equation}

Inserting (\ref{Hrel2}) into (\ref{integrrelation}) leads to
  \begin{equation}\label{integrrelation2}
\int^{\infty}_{x}\frac{dx}{H_{0}\sqrt{\Omega_{m0}}}\sqrt{a(1-\Omega_{DE})}=\frac
{1}{a}\left( \frac{3H^2 \Omega_{DE}-
\sqrt{9H^4 \Omega_{DE}^2 -12c^2 K^2M_p^4} }{2K^2M_p^4}
\right)^{\frac{1}{2}}.
\end{equation}
In the following we  use  $x=\ln a$ as the 
independent variable, and therefore   for a quantity $f$ we acquire 
$\dot{f}=f' H$, 
with 
primes denoting derivatives with respect to $x$. Differentiating 
(\ref{integrrelation2}) in terms of $x$ we   obtain
  \begin{eqnarray}\label{Odediffeq}
&&
\!\!\!\!\!\!\!\!\!\!\!
\Omega_{DE}'=\Omega_{DE}(1-\Omega_{DE})\left 
\{3-\frac{2(\mathcal{A}-2K^{2}M_{p}^{4}\mathcal{B})}{\mathcal{A}}\left 
[1-\sqrt{3}\left (\frac{\Omega_{DE}}{\mathcal{A}\mathcal{B}}\right 
)^{\frac{1}{2}}\right ]\right \},
\end{eqnarray}
with
\begin{eqnarray}\nonumber
\!\!\!\!\!\!\!\!\!\!\!
\mathcal{A}
&=&\frac{3\e^{-3x}H^{2}_{0}\Omega_{m0}\Omega_{DE}}{1-\Omega_{DE}}, 
\\ 
\nonumber
\!\!\!\!\!\!\!\!\!\!\!\!\!\!\!\!\!\!\!\!
\mathcal{B}&=&\frac{\mathcal{A}-\sqrt{\mathcal{A}^{2}-12c^{2}K^{2}M^{4}_{p}}}{
2K^{2}M^{4}_{p}}.
\end{eqnarray}
 
Differential equation (\ref{Odediffeq})  
determines the 
evolution of Kaniadakis holographic dark energy  as 
a function of $x=\ln a$, in the case of flat spatial 
geometry   and for dust matter.
 We mention that in the limit $K\rightarrow 0$ we have 
$\mathcal{B}|_{K\rightarrow 
0}=\frac{3c^{2}}{\mathcal{A}}$, and hence  (\ref{Odediffeq})   recovers   
the corresponding differential equation of usual holographic dark energy  
\cite{Li:2004rb},  
i.e.
$\Omega_{DE}'|_{K\rightarrow 0}= 
\Omega_{DE}(1-\Omega_{DE})\left(1+2\sqrt{\frac{3M_p^2\Omega_{DE}}{3 
c^2 M_p^2}}
\right)$, 
 which, since the $x$-dependence is absent, accepts an 
analytic solution in an implicit form  \cite{Li:2004rb}. 

 We proceed by examining the behavior of the  equation-of-state parameter 
$w_{DE}\equiv p_{DE}/\rho_{DE}$ of
Kaniadakis holographic dark energy. From the conservation of the 
matter sector (\ref{rhoconserv}), and using the two 
Friedmann 
equations (\ref{Fr1b}),(\ref{Fr2b}), we deduce that  the dark energy 
sector is 
conserved too, 
i.e. 
\begin{equation}\label{rhodeconserv}
\dot{\rho}_{DE}+3H\rho_{DE}(1+w_{DE})=0.
\end{equation}
Differentiating (\ref{KHDE2}) gives
$\dot{\rho}_{DE}=2M^{2}_{p}\left (-3c^{2}R^{-4}_{h}+K^{2}M^{4}_{p}\right 
)R_{h}\dot{R}_{h}$.  In this expression we have that $\dot{R}_h=H  R_h-1$, as 
it is   found from (\ref{futrhoriz}),
 where $R_h$ can be further eliminated in terms of $\rho_{DE}$  according to 
(\ref{KHDE2})
as 
\begin{eqnarray}
 R_h=\left 
(\frac{\rho_{DE}-\sqrt{\rho^{2}_{DE}-12c^{2}K^{2}M^{8}_{p}}}{2K^{2}M^{6}_{p}}
\right )^{1/2}\equiv   \mathcal{C}.
\end{eqnarray}
Substituting all
the above into 
(\ref{rhodeconserv}) we acquire 
\begin{eqnarray}
\label{rhodeconserv2}
2M^{2}_{p}(H \mathcal{C}-1 )\left(  
\frac{-3c^{2}+K^{2}M^{4}_{p}\mathcal{C}^{4}}{\mathcal{C}^{3 }}\right 
)+3H\rho_{DE}(1+w_{DE})=0.
\end{eqnarray}
 Therefore, inserting $H$ from (\ref{Hrel2}), and using    definition 
(\ref{ODE}) after some algebra we find
\begin{equation}\label{wDE}
w_{DE}=-1-2\left (\frac{\Omega_{DE}}{3\mathcal{A}^{3}} \right 
)^{\frac{1}{2}}\left 
(\frac{-3c^{2}+K^{2}M^{4}_{p}\mathcal{B}^{2}}{\mathcal{B}^{\frac{3}{2}}}\right 
)\left [-1+\frac{\sqrt{3}}{3}\left (\frac{\mathcal{A}\mathcal{B}}{\Omega_{DE}} 
\right )^{\frac{1}{2}}\right ].
\end{equation}
Hence, $w_{DE}$ as a function of $\ln a$ is known,  as long as 
$\Omega_{DE}$ is known from   (\ref{Odediffeq}). Note that for 
$K\rightarrow 0$ the above expression 
 provides the standard 
holographic 
dark energy result, i.e. 
$w_{DE}|_{K\rightarrow 0}=-\frac{1}{3}-\frac{2}{3}\frac{\sqrt{\Omega_{DE}}}{c}$ 
\cite{Wang:2016och}, as expected. Additionally,
we mention that in general   $w_{DE}$ can be either quintessence-like or 
phantom-like,    which 
is an advantage revealing the rich capabilities of the scenario at hand.

Lastly, for convenience we can  introduce the   deceleration 
parameter  
  \begin{equation}
  \label{qdeccel}
q\equiv-1-\frac{\dot{H}}{H^2}=\frac{1}{2}+\frac{3}{2}\left(w_m\Omega_m+w_{DE}
\Omega_{DE}
  \right),
\end{equation}
which  in the case of dust matter   is straightforwardly known as 
long as 
$\Omega_{DE}$ (and thus $w_{DE}$ from (\ref{wDE})) is known.

We close this section by discussing    the relation of Kaniadakis entropy with 
other 
extended  entropies, and in particular with Tsallis one. As it is known,
the non-extensive Tsallis entropy $S^{T}_q$, with $q$ the parameter which 
quantifies the deviation from Bekenstein-Hawking entropy 
\cite{Tsallis:1987eu,Tsallis:2012js}, is related to Kaniadakis one through
 \cite{Abreu:2017hiy,Moradpour:2020dfm,Nunes:2015xsa}
\begin{equation}
S_{K} =\frac{S^{T}_{1+K}+S^{T}_{1-K}}{2}.
 \end{equation}
Concerning the other recently proposed generalized  entropy  by Barrow, namely
$S^{B}_\Delta$,  which arises from  the intricate structure of the black-hole
surface due to  quantum-gravitational 
effects, with    $\Delta$   the parameter that 
quantifies the deviation from usual entropy \cite{Barrow:2020tzx}, we mention 
that 
although mathematically one can extract the relation  $
S_{K} =\frac{S^{B}_{\Delta}+S^{B}_{-\Delta}}{2}$,  it cannot have a 
physical application since in Barrow entropy $0\leq\Delta\leq1$.

\section{Cosmological evolution}
\label{Cosmologicalevolution}

In the previous section we formulated Kaniadakis holographic dark energy, and 
we provided the equations that determine the evolution of the corresponding 
dark 
energy density, equation-of-state and deceleration parameters. Hence, we can 
now proceed  to a detailed investigation of the resulting cosmological 
behavior.  Since  equation  (\ref{Odediffeq})  
can be solved analytically   only for  $K=0$, in the general case we should 
resort to numerical elaboration. As long as we have the solution for 
$\Omega_{DE}(x)$  we can obtain its behavior in terms of the redshift $z$ 
through the simple relation $x\equiv\ln a=-\ln(1+z)$.
Finally, we mention   that Kaniadakis entropy  is an even 
function, namely $S_{K}=S_{-K}$, and that is why all the 
  above expressions of Kaniadakis   holographic dark energy 
depend only on $K^2$. Thus, in the following we 
focus on the  $K\geq0$ region.

\begin{figure}[!]
\centering
\includegraphics[width=7.cm]{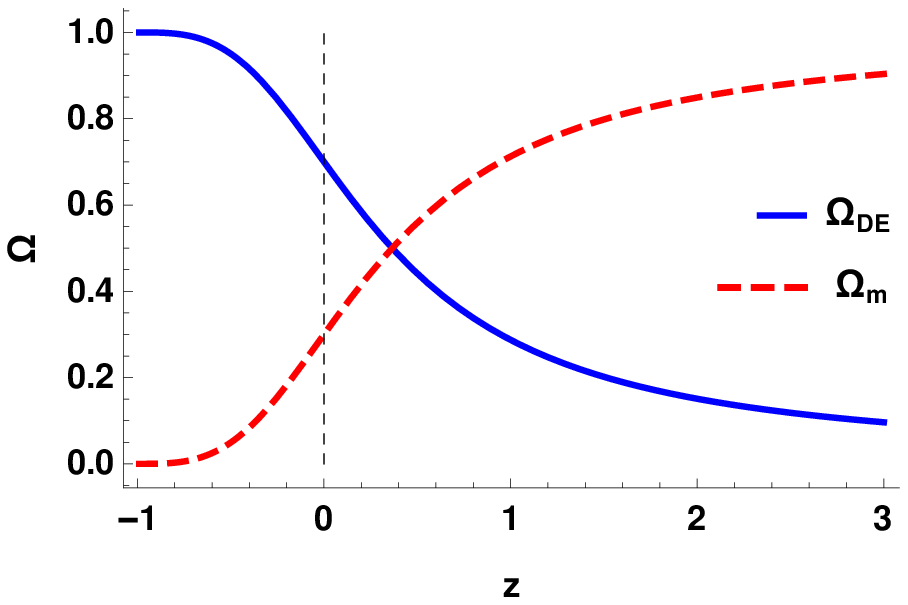}    \\                           
       \includegraphics[width=7.cm]{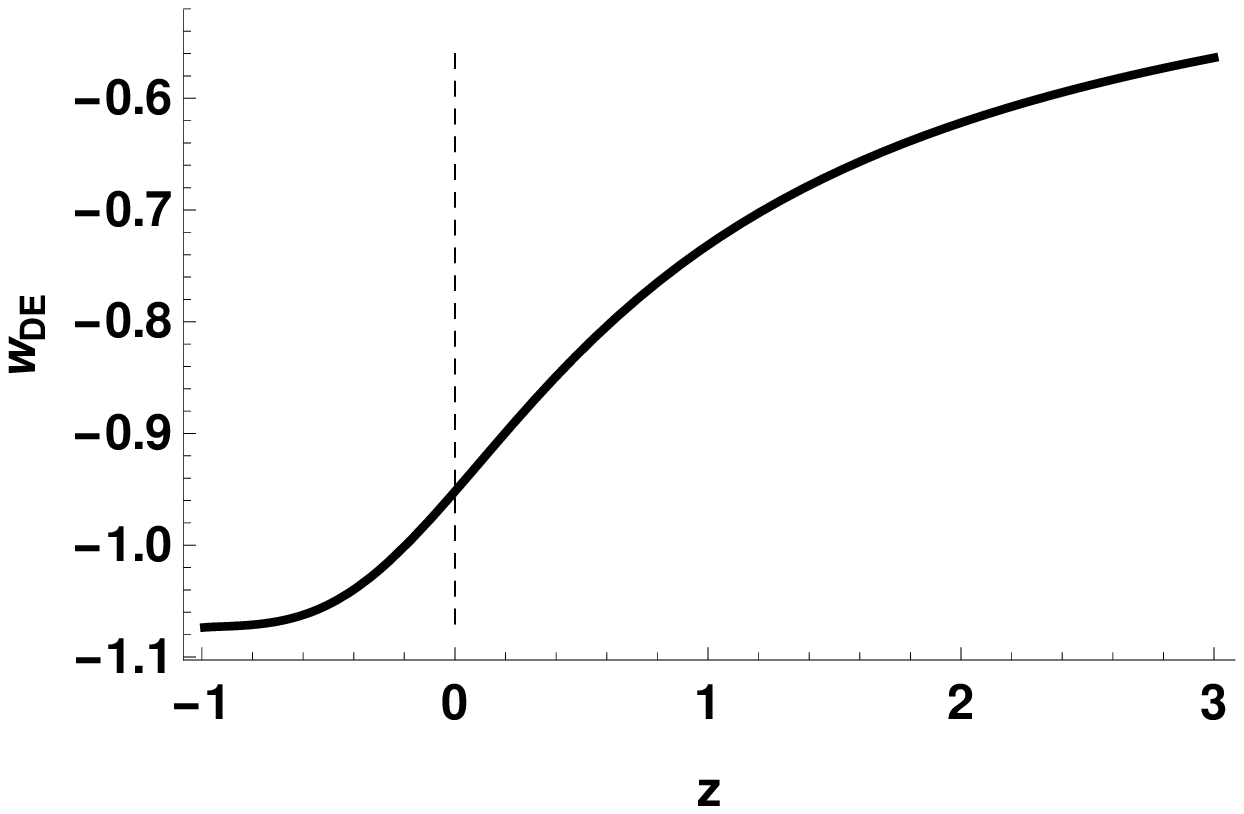} \\
\includegraphics[width=7.cm]{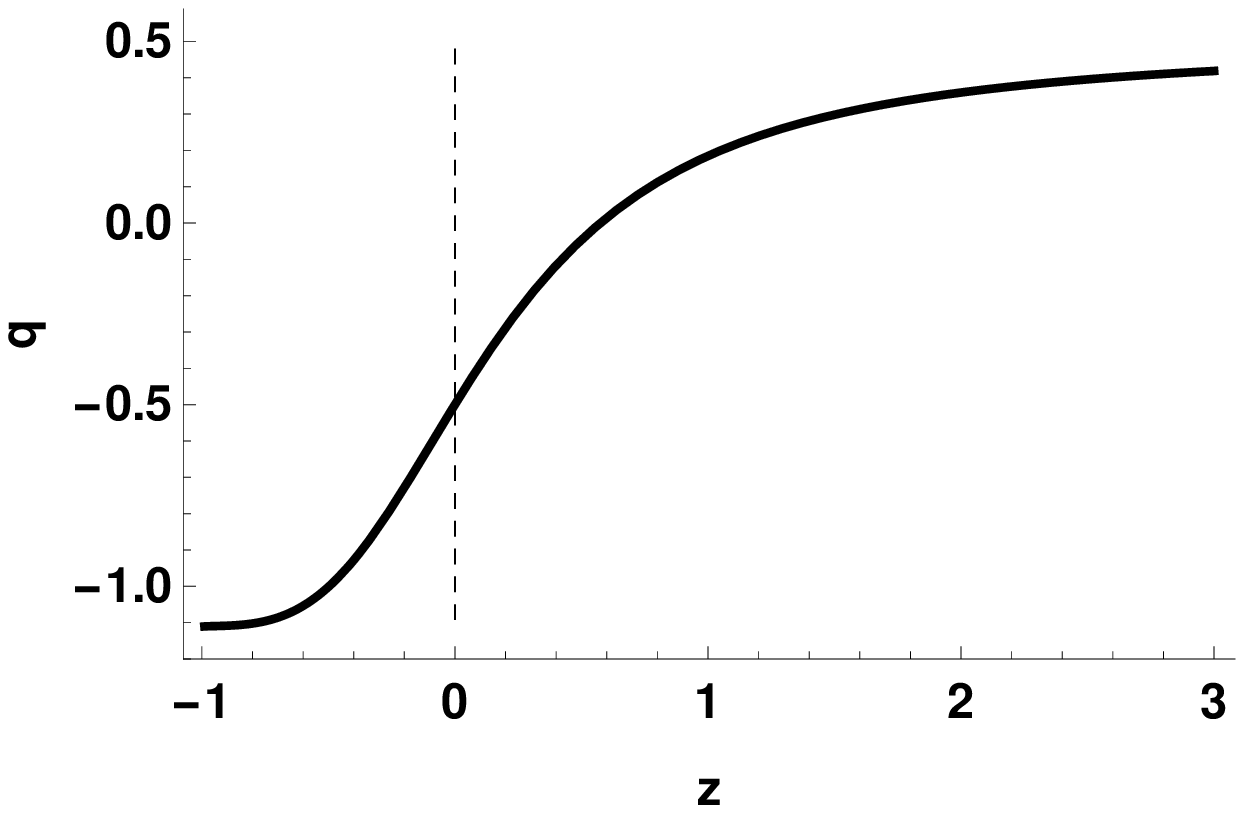}
\caption{\it{ {\bf{Upper graph}}: The    Kaniadakis holographic dark energy   
 density   parameter $\Omega_{DE}$ (blue-solid)  and   the matter density
parameter $\Omega_{m}$ (red-dashed), as a function of the redshift $z$, for
$K=0.1$ and $c=0.9$. 
 {\bf{ Middle graph}}: The    corresponding dark-energy equation-of-state 
parameter $w_{DE}$. {\bf{Lower graph}}:  The   
corresponding   deceleration parameter $q$. In all graphs we have set 
 $\Omega_{DE}(x=-\ln(1+z)=0)\equiv\Omega_{DE0}\approx0.7$    in 
agreement with observations, and for convenience we have added a vertical 
dotted line marking the present time $z=0$.
 }}
\label{HDEOmegas}
\end{figure}

We solve equation   (\ref{Odediffeq}) numerically, imposing   
$\Omega_{DE}(x=-\ln(1+z)=0)\equiv\Omega_{DE0}\approx0.7$ and therefore 
$\Omega_m(x=-\ln(1+z)=0)\equiv\Omega_{m0}\approx0.3$ in agreement with
observations 
\cite{Planck:2018vyg}.  In the upper graph of Fig. \ref{HDEOmegas} we depict 
the evolution of  the dark energy   and matter
 density   parameters    in terms of the
redshift. Additionally, in the middle graph we present  the 
corresponding behavior of the dark-energy equation-of-state parameter as it 
arises from  (\ref{wDE}). Finally,  
in the 
lower graph we show the deceleration parameter as it is given from 
(\ref{qdeccel}).
We mention that for reader's convenience
 we have extended the 
evolution up to the far future, namely for   $z\rightarrow-1$.   

As we observe,  the scenario at hand can provide the required thermal 
history of the universe, i.e. the sequence of matter and dark energy epochs, 
and the universe results asymptotically to a complete dark-energy dominated 
phase. Moreover, from the middle graph of Fig. \ref{HDEOmegas} we can see 
that the value of $w_{DE}$ at present is around $-1$ in agreement with  
observational data. Note that in this specific example $w_{DE}$  in the future 
enters slightly inside the phantom 
regime, 
which as mentioned above is allowed by 
 (\ref{wDE}) and shows the capabilities of the model.
Finally, from the lower graph of Fig. \ref{HDEOmegas} we deduce that the 
transition from deceleration to acceleration is realized at  $z\approx 0.6$, in 
agreement with observations.

\begin{figure}[!h]
\centering
\includegraphics[width=10.5cm]{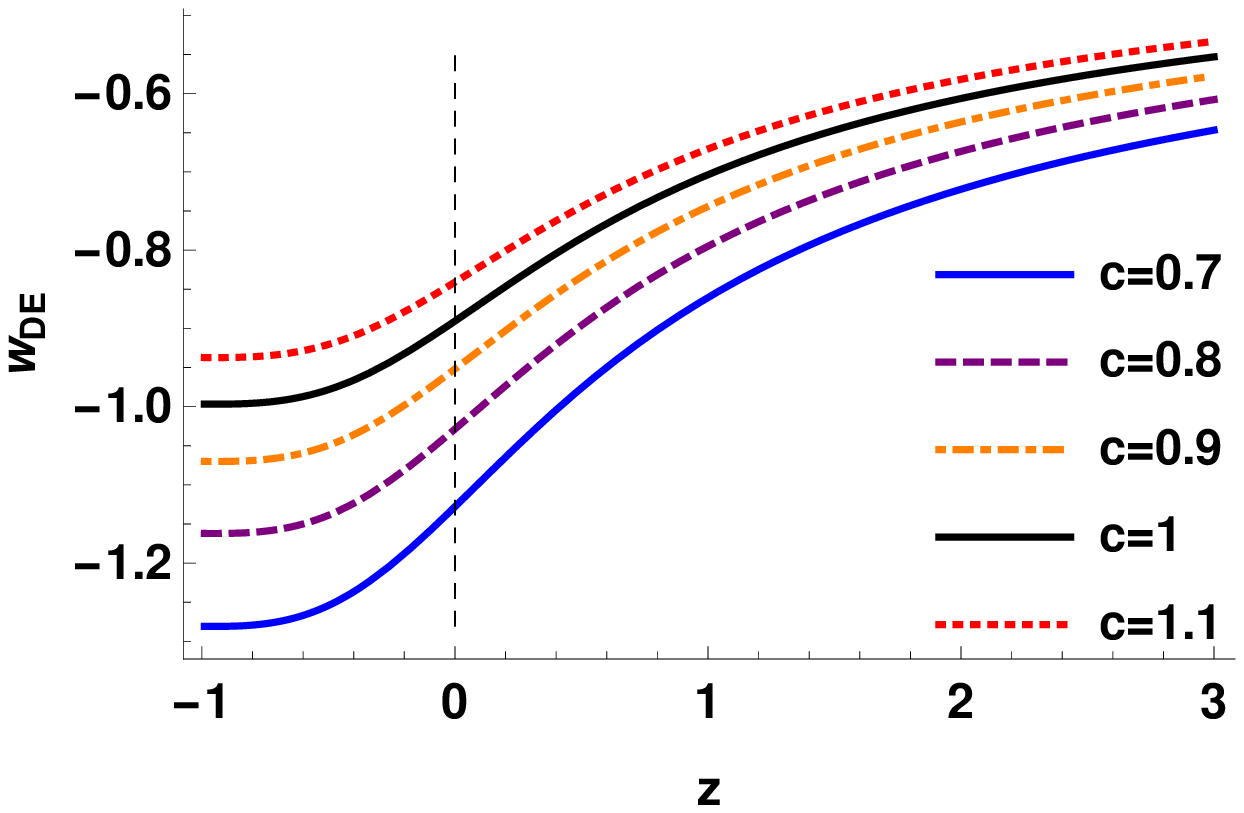}                             
  \caption{\it{ The redshift-evolution of the equation-of-state parameter 
$w_{DE}$ of Kaniadakis
holographic dark energy,    for fixed $K=0.1$ and various values of $c$.
We have imposed 
$\Omega_{DE0}\approx0.7$  and   we have added a vertical 
dotted line marking the present time $z=0$.
}}
\label{HDEmultiwdec}
\end{figure} 
 
\begin{figure}[!h]
\centering
\includegraphics[width=10.5cm]{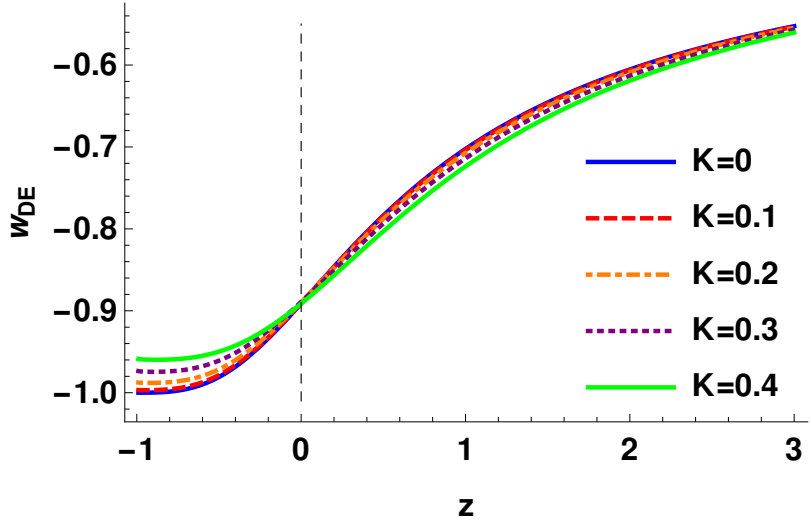}                             
   \caption{\it{ The redshift-evolution of the equation-of-state parameter 
$w_{DE}$ of Kaniadakis
holographic dark energy,    for fixed $c=1$ and various values of $K$.
We have imposed 
$\Omega_{DE0}\approx0.7$  we have added a vertical 
dotted line marking the present time $z=0$.
}}
\label{HDEmultiwdeK}
\end{figure}

Let us now study the effect of the model parameters $c$ and $K$ on the    
dark-energy  equation-of-state  parameter $w_{DE}$. In Fig. \ref{HDEmultiwdec} 
we depict $w_{DE}(z)$ for fixed $K=0.1$ and various values of $c$. As we can 
see, with  $c$  decreasing $w_{DE}(z)$, as well as its present value 
$w_{DE}(z=0)$, acquire algebraically  lower values, experiencing the 
phantom-divide crossing during the evolution. Note that for $c<0.9$ 
the value of $w_{DE}(z=0)$ lies in the phantom regime.
Furthermore, in 
  Fig. \ref{HDEmultiwdeK} we present $w_{DE}(z)$ for fixed $c=1$ and 
various values of $K$. Here we observe the interesting behavior that for 
increasing $K$, at earlier times  $w_{DE}$ slightly decreases,   in future 
times in increases, however at times around the present ones it  remains almost 
unaltered. Concerning the asymptotic value of $w_{DE}$ in the far future, 
namely for $z\rightarrow-1$, as can be deduced from the figures, as well as 
form (\ref{wDE}), it depends on the combination of $K$ and $c$.
In summary, we can see that the scenario of Kaniadakis holographic dark energy 
can lead to very interesting cosmological phenomenology, in which $w_{DE}$ can 
be    quintessence-like, phantom-like, or cross  the phantom divide 
  before or after the present time.  Note that these 
results are in agreement with other researches on Kaniadakis holographic dark 
energy, that appeared after the present work, especially with  observational 
 confrontation  
\cite{Hernandez-Almada:2021rjs,Lymperis:2021qty,
Luciano:2022knb,Hernandez-Almada:2021aiw, Ghaffari:2021xja, Sadeghi:2022fow}.

\section{Conclusions}
\label{Conclusions}
 
 In this work we formulated a holographic dark energy scenario based on 
Kaniadakis  entropy. The latter is a generalization of  Boltzmann-Gibbs  
entropy, arising form a coherent relativistic statistical 
theory and  characterized by a single parameter $K$ that quantifies the
deviations from standard expressions. Hence, by applying the usual steps of 
holographic dark energy, imposing the future event 
horizon as the  Infrared  cutoff, and using Kaniadakis extended entropy, we 
obtained  
Kaniadakis holographic dark energy in a consistent way, namely a one-parameter 
extension of usual holographic dark energy, possessing it as a particular 
limit, namely for $K\rightarrow0$.

In order to investigate the cosmological application of Kaniadakis holographic 
dark energy we extracted the  differential equation that determines the 
evolution of the effective dark energy density 
parameter $\Omega_{DE}$. Moreover, we provided   analytical expressions for 
the  corresponding equation-of-state parameter $w_{DE}$, as well as for the 
deceleration parameter.

The scenario of Kaniadakis holographic dark energy proves to lead to 
interesting cosmological behavior. In particular, the universe exhibits the 
standard thermal history, i.e. the sequence of matter and dark-energy eras, 
while  the transition to acceleration takes place at $z\approx0.6$.
Concerning the dark-energy equation-of-state parameter we saw that it can have 
a rich behavior, being quintessence-like, phantom-like, or experience the 
phantom-divide crossing in the past or in the future, depending on the values 
of 
the two model parameters $c$ and $K$. In particular, for fixed $K$ decreasing 
$c$ leads to algebraically smaller  $w_{DE}$ values, while for fixed $c$ by
increasing $K$  we acquire smaller $w_{DE}$  values  at higher redshifts, 
larger $w_{DE}$  values  in the future, and almost   
unaltered values at present. Finally, in the far future dark energy 
dominates completely, and the asymptotic $w_{DE}$ value depends on 
 $c$ and $K$.

 We comment here that, as we mentioned in the Introduction, it has been 
recently shown that holographic dark energy constructions may alleviate the  
$H_0$ tension (see the corresponding sections in the recent review 
\cite{Abdalla:2022yfr}), and the reason is that they may lead to $w_{DE}<-1$ 
which seems to be a requirement if one desires to provide a solution based on 
  late-time modifications \cite{Colgain:2021beg,Vagnozzi:2019ezj}. As we 
saw, the scenario at hand 
 can fulfill  this requirement and that is why it is a candidate to be able to 
alleviate the $H_0$ 
tension too.  Definitely the phantom regime may have potential disadvantages, 
however this is not in general the case in  
models which present phantom behavior in an effective way, 
  among which is holographic dark energy  \cite{Wang:2016och}.

In conclusion, Kaniadakis holographic dark energy exhibits richer and more 
interesting behavior in comparison to usual holographic dark energy. 
Additionally, due to the consistent formulation, it possesses the latter as a 
limiting sub-case. Definitely, before one considers it as a successful 
candidate for the description of dark energy, there are necessary 
investigations that should be performed. In particular, one should confront the 
scenario with observational data from  Supernova type Ia (SNIa),   
Baryon Acoustic Oscillation (BAO), Cosmic Microwave Background (CMB), 
and Hubble parameter observations, and extract constraints on the model 
parameters. Additionally, one should analyze in detail the phase-space 
behavior, in order to examine the global dynamics and the asymptotic, late-time 
evolution of the scenario. These   investigations will be performed in separate 
 projects.

\begin{acknowledgments} 
The authors would like to acknowledge the contribution of the COST Action 
CA18108 ``Quantum Gravity Phenomenology in the multi-messenger approach''.
  The work is partially supported by the Ministry of Education and Science of 
the Republic of Kazakhstan, Grant AP08856912.
 
\end{acknowledgments}



\end{document}